\documentclass[twocolumn]{aastex63}

\graphicspath{{./}{figures/}}
\usepackage{pifont}
\usepackage{multirow}
\usepackage{amsmath}
\usepackage{xcolor}

\begin{document}

\title{Formation of GW230529 from Isolated Binary Evolution}

\author[0000-0002-9195-4904]{Jin-Ping Zhu}
\altaffiliation{These authors contributed equally to this work}
\affiliation{School of Physics and Astronomy, Monash University, Clayton Victoria 3800, Australia}
\affiliation{OzGrav: The ARC Centre of Excellence for Gravitational Wave Discovery, Australia}

\author[0000-0002-6442-7850]{Rui-Chong Hu}
\altaffiliation{These authors contributed equally to this work}
\affiliation{Nevada Center for Astrophysics, University of Nevada, Las Vegas, NV 89154, USA}
\affiliation{Department of Physics and Astronomy, University of Nevada, Las Vegas, NV 89154, USA}

\author[0000-0001-7402-4927]{Yacheng Kang}
\affiliation{Department of Astronomy, School of Physics, Peking University, Beijing 100871, China}
\affiliation{Kavli Institute for Astronomy and Astrophysics, Peking University, Beijing 100871, China}

\author[0000-0002-9725-2524]{Bing Zhang}
\affiliation{Nevada Center for Astrophysics, University of Nevada, Las Vegas, NV 89154, USA}
\affiliation{Department of Physics and Astronomy, University of Nevada, Las Vegas, NV 89154, USA}

\author[0000-0002-4534-0485]{Hui Tong}
\affiliation{School of Physics and Astronomy, Monash University, Clayton Victoria 3800, Australia}
\affiliation{OzGrav: The ARC Centre of Excellence for Gravitational Wave Discovery, Australia}

\author[0000-0002-1334-8853]{Lijing Shao}
\affiliation{Kavli Institute for Astronomy and Astrophysics, Peking University, Beijing 100871, China}
\affiliation{National Astronomical Observatories, Chinese Academy of Sciences, Beijing 100012, China}

\author[0000-0002-2956-8367]{Ying Qin}
\affiliation{Department of Physics, Anhui Normal University, Wuhu, Anhui 241002, China}
\affiliation{Purple Mountain Observatory, Chinese Academy of Sciences, Nanjing 210023, China}

\correspondingauthor{Jin-Ping Zhu, Rui-Chong Hu, Bing Zhang}
\email{jin-ping.zhu@monash.edu, ruichong.hu@unlv.edu, bing.zhang@unlv.edu}

\begin{abstract}

In this paper, we explore the formation of the mass-gap black hole-neutron star (mgBHNS) merger detected in gravitational wave (GW) event, i.e., GW230529, from the isolated binary evolution channel, and study potential signatures of its electromagnetic counterparts. {By adopting the `delayed' supernova prescription and reasonable model realizations, our population synthesis simulation results can simultaneously match the rate densities of mgBHNS and total BHNS mergers inferred from the population analyses, along with the population distribution of the BH mass in BHNS mergers reported by the LIGO-Virgo-KAGRA Collaboration. Because GW230529 contributes significantly to the inferred mgBHNS rate densities, we suggest that GW230529 can be explained through the isolated binary evolution channel.} Considering the AP4 (DD2) equation of state, the probability that GW230529 can make tidal disruption is $12.8\%$ ($63.2\%$). If GW230529 is a disrupted event, its kilonova peak apparent magnitude is predicted $\sim23-24\,{\rm{mag}}$, and hence, can be detected by the present survey projects and LSST. {Since GW230529 could be an off-axis event inferred from the GW observation, its associated gamma-ray burst (GRB) might be too dim to be observed by $\gamma$-ray detectors, interpreting the lack of GRB observations.} {Our study suggests} the existence of mgBHNS mergers formed through the isolated binary evolution channel {due to the discovery of GW230529}, {indicating} that BHNS mergers are still likely to be multimessenger sources that emit GWs, GRBs, and kilonovae. Although mgBHNS mergers account for {$\sim50\%$} cosmological BHNS population, we find that $\gtrsim90\%$ disrupted BHNS mergers are expected to originate from mgBHNS mergers.

\end{abstract}


\section{Introduction} \label{sec:intro}

Black hole--neutron star (BHNS) mergers are prime search targets for the ground-based gravitational-wave (GW) detectors, including LIGO \citep{LIGO2015}, Virgo \citep{Acernese2015}, and KAGRA \citep{Aso2013}. Until the end of the GW third observing run (O3), the first two BHNS mergers were identified by the LIGO-Virgo-KAGRA (LVK) Collaboration \citep{Abbott2021Observation}, with GW200105\footnote{{GW200105 was reported to be a marginal event with a probability of astrophysical origin $p_{\rm astro}<0.5$ in GWTC-3 \citep{Abbott2023GWTC3}.}} being a merger between a $8.9^{+1.2}_{-1.5}M_\odot$ BH and a $1.9^{+0.3}_{-0.2}M_\odot$ NS, and GW200115 being a merger between a $5.7^{+1.8}_{-2.1}M_\odot$ BH and a $1.5^{+0.7}_{-0.3}M_\odot$ NS {(all measurements quoted at the 90\% credible level)}. During O3, the LVK Collaboration reported three additional marginal BHNS candidates \citep{Abbott2021GWTC2,Abbott2021GWTC21,Abbott2023GWTC3}, with GW190426\_152155 involving a $5.7^{+3.9}_{-2.3}M_\odot$ BH and a $1.5^{+0.8}_{-0.5}M_\odot$ NS, GW190917\_114630 involving a $9.7^{+3.4}_{-3.9}M_\odot$ BH and a $2.1^{+1.1}_{-0.4}M_\odot$ NS, as well as GW191219\_163120 involving a $31.1^{+2.2}_{-2.8}M_\odot$ BH and a $1.17^{+0.07}_{-0.06}M_\odot$ NS. GWTC-3 also includes a puzzling GW event, GW190814, composed of a $22.2-24.3\,M_\odot$ primary BH and a $2.50-2.67\,M_\odot$ secondary compact object with unclear origin \citep{Abbott2020GW190814}. The LVK Collaboration also discovered a similar, but marginal GW candidate, namely GW200210\_092254 \citep{Abbott2023GWTC3}, with component masses inferred to be $24.1^{+7.5}_{-4.6}M_\odot$ and $2.83^{+0.47}_{-0.42}M_\odot$, respectively. 

The X-ray and radio observations of Galactic pulsars implied a likely NS maximum mass of $\sim2-2.3\,M_\odot$ \citep[e.g.,][]{Antoniadis2013,Alsing2018,Romani2022} while Galactic BHs in X-ray binaries were inferred to have a lower boundary close to $\sim5\,M_\odot$ \citep{Bailyn1998,Ozel2010,Farr2011}, leading to the conjecture of the presence of a mass gap between the heaviest NSs and lightest BHs. However, recent electromagnetic (EM) observations of non-interacting binary systems \citep{Thompson2019,Rivinius2020,Andrews2022} and gravitational microlensing events \citep{Wyrzykowski2020} indicated that the mass gap might be partially polluted. The population properties of O3 BHNS candidates suggested a relative dearth of events with masses in the mass gap \citep{Zhu2022Population,Ye2022,Biscoveanu2023}, which was also supported by the population study on merging compact binaries in GWTC-3 \citep{Abbott2023Population}. Whether there is a mass gap or not can shed light on supernova (SN) mechanisms for the formation of NSs and BHs, with the rapid model giving rise to the mass gap, while the delayed model does not \citep{Fryer2012}. Population synthesis simulations revealed that BHs in $\sim30–80\%$ BHNS mergers can have a mass in the mass gap by considering the delayed SN model \citep{Shao2021,Drozda2022}. It is expected that future detection of merging mass-gap BHNS (mgBHNS) binaries through GW observations can give a better constraint on the SN mechanisms.

Binary NSs (BNSs) and BHNSs have long been proposed to be progenitors of some fast-evolving EM transients, including gamma-ray bursts \citep[GRBs;][]{Paczynski1986,Paczynski1991,Eichler1989,Narayan1992,Zhang2018,Gottlieb2023} and kilonovae \citep{Li1998,Metzger2010}. While BNS mergers were confirmed as the origin of GRBs and kilonovae thanks to the associations between GW170817 \citep{Abbott2017GW170817}, GRB170817A \citep{Abbott2017Gravitational,Goldstein2017,Savchenko2017,ZhangBB2018}, and AT2017gfo \citep[e.g.,][]{Abbott2017Multimessenger,Arcavi2017,Coulter2017,Drout2017,Evans2017,Kasen2017,Kasliwal2017,Kilpatrick2017,pian2017,Smartt2017}, the joint observations of the O3 BHNS GW candidates and their associated EM counterparts, especially for kilonova emissions, were absent. One possible reason for the lack of kilonova detection following O3 BHNS GW signals could be the challenge of rapidly achieving full distance and volumetric coverage for the probability maps of the LVK Collaboration within the short kilonova duration by current survey projects \citep[e.g.,][]{Coughlin2020,Kasliwal2020,Gompertz2020,Anand2021}. Unlike BNS mergers, which typically eject a certain amount of materials to produce EM counterparts, the NS components in some merging BHNS binaries may directly plunge into their BH companions without generating any observable EM signals. NS tidal disruption tends to occur if the BHNS binaries have a low-mass BH with a high orbital aligned spin and a low-mass NS with a stiff EoS \citep{Shibata2011,Foucart2012,Kyutoku2013,Kyutoku2015,Kawaguchi2016,Foucart2018,Zhu2020,Zhu2021Kilonova,Hayashi2021,Sarin2022,Clarke2023}. More specifically, since primary BHs (produced from initially more massive star) formed through the classical CE scenario typically possess near-zero aligned spins \citep{Qin2018,Fuller2019,Belczynski2020}, tidal disruptions of most cosmological BHNS mergers are expected to occur if the BHs have a mass of $\lesssim6-7\,M_\odot$ and the NSs have a mass of $\lesssim1.5\,M_\odot$ \citep[e.g.,][]{Zhu2021No,Zhu2022Population}. If the mass gap does exist, the mass space that allows NS tidal disruption and produces bright EM signals could be limited. Because O3 BHNS candidates have BH masses of $\gtrsim5\,M_\odot$ inferred from GW observations, their posterior mass distributions mostly lie outside the tidal disruption mass range \citep{Abbott2021Observation,Zhu2021No,Zhu2022Population,DOrazio2022}. Thus, it is likely that the EM counterparts associated with these BHNS candidates were intrinsically missing. Additionally, although a few O3 BHNS candidates still have low probabilities of undergoing tidal disruption and producing kilonova emissions, the brightness of these kilonovae might be too dim to be detected by current survey telescopes due to their remote distance from us \citep{Zhu2021No}.

The discovery of mgBHNS mergers through GW observations can provide an opportunity to constrain SN mechanisms and isolated formation channel of compact binaries. Furthermore, mgBHNS mergers are more likely to have tidal disruption and produce observable EM counterparts. Multimessenger observations of BHNS mergers between GWs and EMs can be more easily achieved if mgBHNS mergers exist. Most recently, the LVK Collaboration reported the first mgBHNS merger, GW230529, detected in the first part of the fourth observing run \citep{Abbott2024Observation}. {By adopting the combined posterior results inferred by the low-spin prior of the secondary component} \citep[{i.e., the posterior sample of \texttt{Combined\_PHM\_lowSecondarySpin};}][]{Abbott2024Observation},  the primary BH mass of GW230529 is $M_{\rm BH}=3.6^{+0.7}_{-1.2}M_\odot$ {with inferred $90\%$ credible region nearly occupying the mass gap}, while its secondary NS mass is $M_{\rm NS}=1.43^{+0.59}_{-0.19}M_\odot$. The effective inspiral-spin and effective precessing-spin of GW230529 are $\chi_{\rm eff} = -0.10^{+0.10}_{-0.17}$ and $\chi_{\rm p} = 0.40^{+0.37}_{-0.34}$, respectively. The source has a redshift of $z=0.043^{+0.023}_{-0.021}$, corresponding to a luminosity distance of $D_{\rm L}=197^{+107}_{-96}{\rm Mpc}$. {Based on the population posterior of currently detected BHNS mergers using the NSBH-\textsc{pop} model\footnote{{The NSBH-\textsc{pop} model is a parametric model to constrain the population properties of the BH mass, NS mass, and BHNS rate density {as a function of BH and NS masses. The BHNS rate density can be calculated by setting a specific mass range.} See Appendix H.3 in \cite{Abbott2024Observation}} for more details of this population model.} reported {in Figure 4} of \cite{Abbott2024Observation}, we find the rate densities of mgBHNS mergers with a primary BH mass $\lesssim5\,M_\odot$ and total BHNS mergers to be $R_{\rm mgBHNS}=18^{+58}_{-16}\,{\rm Gpc}^{-3}{\rm yr}^{-1}$ and $R_{\rm BHNS}=40^{+77}_{-29}\,{\rm Gpc}^{-3}{\rm yr}^{-1}$, respectively. Without the inclusion of GW230529, these rate densities are changed to  $R_{\rm mgBHNS}=0^{+10}_{-0}\,{\rm Gpc}^{-3}{\rm yr}^{-1}$ and $R_{\rm BHNS}=18^{+41}_{-15}\,{\rm Gpc}^{-3}{\rm yr}^{-1}$. Thus, GW230529 significantly increases the inferred rate of mgBHNS mergers.} 

In this paper, we explore the formation of GW230529 through the isolated binary evolution channel and study the properties of its associated EM counterparts. {Hereafter, the combined posterior sample inferred by the default low-spin prior of the secondary component released by the LVK Collaboration is employed for our studies on individual GW sources, including GW200105, GW200115, and GW230529 \citep{Abbott2021Observation, Abbott2024Observation}.}

\section{Population Synthesis of BHNS Mergers and Formation of GW230529}

\subsection{Method}

In this section, we employ the rapid binary population synthesis code \texttt{COMPAS} {\citep[version 02.39.00;][]{Stevenson2017,Vigna2018,Neijssel2019,TeamCOMPAS2022}} to explore the formation of GW230529 through the isolated binary evolution channel. To determine compact object masses during core-collapse SNe (CCSNe) of stars, the `delayed' SN prescription \citep{Fryer2012} is adopted, which allows for the production of mass-gap BHs \citep[see {also the stochastic recipe developed by}][]{Mandel2020}. {We generate and evolve binary systems according to the fiducial population synthesis model described in Table 1 of \cite{Broekgaarden2021}, with the maximum NS mass changed to $M_{\rm NS,max} = 2.2\,M_\odot$.} {We refer readers to Table \ref{tab:COMPAS} in Appendix for further details on other initial conditions, parameter settings, and simulation settings we used. }

We explore the variations of different common-envelope (CE) efficiencies ($\alpha_{\mathrm{CE}} = 1,\,2,\,5,\,10$) and {the velocity dispersion of CCSNe natal kicks \citep[$\sigma_{\mathrm{CCSN}}=100,\,265\,{\rm km}\,{\rm s}^{-1}$;][]{Hobbs2005} in the formation of BHNS mergers.} The naming convention of our population synthesis models follows the pattern $\alpha X \sigma Y$, where $X$ and $Y$ are the corresponding values of $\alpha_{\rm CE}$ and {$\sigma_{\rm CCSN}$.} {Our population synthesis simulations include 8 models, each evolving $10^6$ binaries across 10 metallicity bins, totaling $10^7$ binaries.} We select BHNS merger systems if the final primary mass $M_1>M_{\rm NS,max}$ and secondary mass $M_2<M_{\rm NS,max}$, {while we identify mgBHNS merger systems by further requiring the final primary mass $M_{\rm NS,max}<M_1<5\,M_\odot$. Since $\chi_{\rm eff}$ and $\chi_{\rm p}$ of GW NSBH mergers were measured to potentially have low distributions, consistent with those of BHNS mergers originating from the isolated binary evolution \citep[e.g.,][]{Zhu2022Population}, we do not consider spin parameters in our selection criteria.} {Using Equation (2) in \cite{Broekgaarden2021}, the redshift-dependent BHNS rate densities are calculated based on the metallicity-specific star formation rate density model, which combines the \cite{Madau2014} star formation rate density with the \cite{Panter2004} galaxy mass function and the \cite{Langer2006} mass-metallicity relation. The local BHNS rate densities are obtained by considering simulated events within a redshift of 0.25.}

\subsection{Results} \label{sec:Results}

\begin{deluxetable}{lccc}[tpb]
\tablecaption{Merger Rate Density of Population Synthesis Simulation} \label{tab:EventRate}
\tablehead{
 \colhead{Model} &
 \colhead{{$\mathcal{R}_{\rm mgBHNS}$}} &
 \colhead{$\mathcal{R}_{\rm BHNS}$} &
 \colhead{CI} \\
 & \colhead{(${\rm Gpc}^{-3}{\rm yr}^{-1}$)} &\colhead{(${\rm Gpc}^{-3}{\rm yr}^{-1}$)} &
}
\startdata
$\alpha1\sigma100$ & $57^{+5}_{-5}$ & $149^{+8}_{-9}$ & $0.90^{+0.02}_{-0.03}$  \\
$\alpha2\sigma100$ & $94^{+8}_{-9}$ & $153^{+10}_{-10}$ & $0.84^{+0.05}_{-0.07}$  \\
$\alpha5\sigma100$ & $74^{+9}_{-7}$ & $86^{+9}_{-8}$ &  $0.47^{+0.17}_{-0.10}$ \\
$\alpha10\sigma100$ & $58^{+7}_{-6}$ & $67^{+7}_{-6}$ & $0.42^{+0.21}_{-0.17}$ \\
$\alpha1\sigma265$ & $9^{+2}_{-2}$ & $39^{+4}_{-4}$ & $0.59^{+0.13}_{-0.59}$ \\
$\alpha2\sigma265$ & $17^{+3}_{-3}$ & $37^{+5}_{-4}$ & $0.15^{+0.15}_{-0.07}$ \\
$\alpha5\sigma265$ & $22^{+4}_{-4}$ & $28^{+5}_{-4}$ & $0.37^{+0.36}_{-0.25}$ \\
$\alpha10\sigma265$ & $16^{+3}_{-2}$ & $25^{+4}_{-4}$ & $0.34^{+0.26}_{-0.15}$
\enddata
\tablecomments{The columns are (1) the population synthesis model; (2) the simulated local merger rate density of {mgBHNS} events in ${\rm Gpc}^{-3}\,{\rm yr}^{-1}$; (3) the simulated local merger rate density of total BHNS mergers in ${\rm Gpc}^{-3}\,{\rm yr}^{-1}$; (4) the credible interval for our simulated population synthesis rate densities to fall within the posterior distribution of rate densities obtained from the population analyses reported by the LVK Collaboration. {The simulated median values of $\mathcal{R}_{\rm mgBHNS}$ and $\mathcal{R}_{\rm BHNS}$, along with their 90\% credible intervals, are obtained via the method of bootstrapping.} }
\end{deluxetable}

\begin{figure}[t]
\centering
\includegraphics[width=1\linewidth, trim = 33 10 82 32, clip]{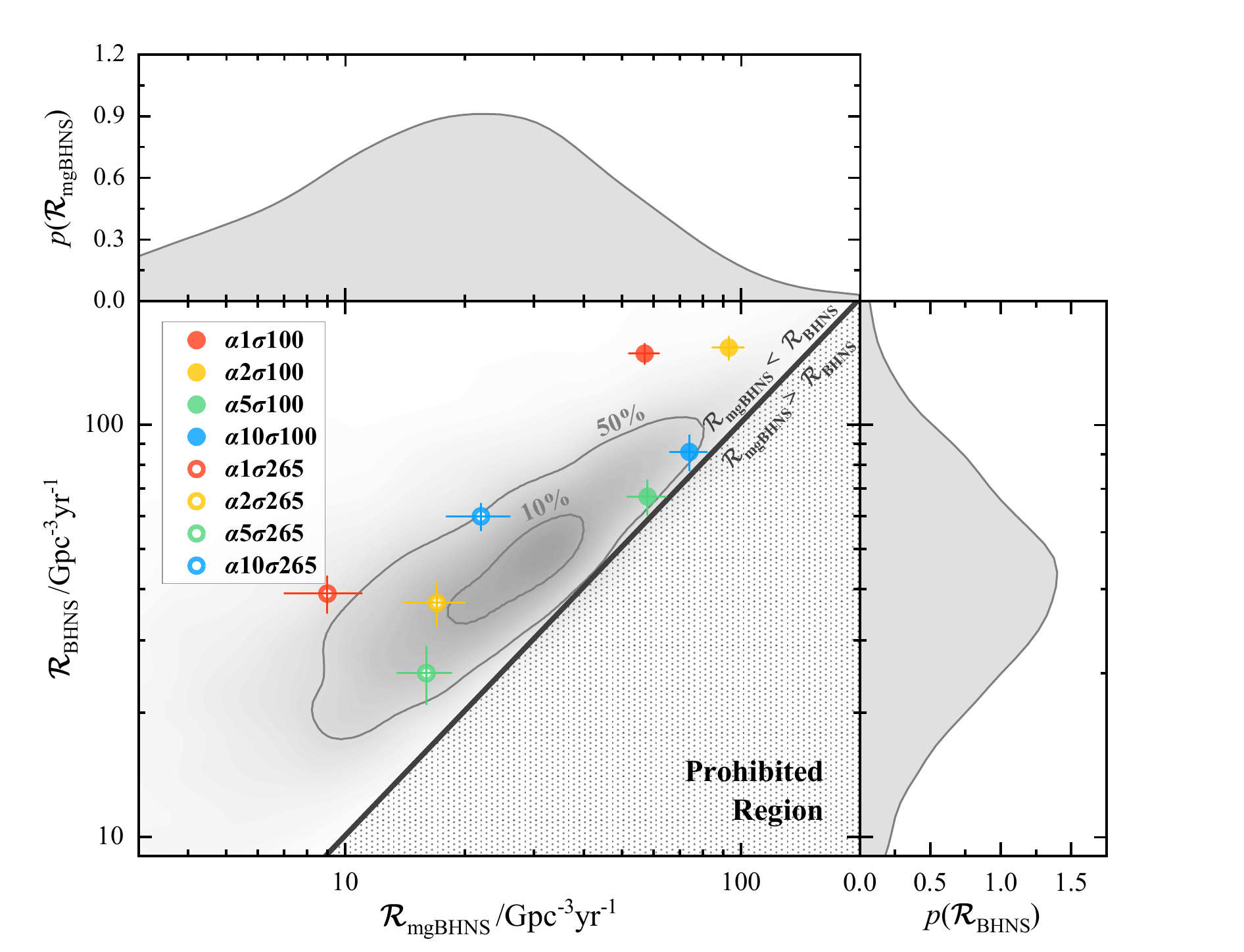}
\caption{Simulated local rate density of {mgBHNS} and BHNS mergers. The red, orange, green, and blue points correspond to different envelope efficiencies of $\alpha_{\rm CE}=1,\,2,\,5,\,{\rm and}\,10$, respectively. The solid and hollow points represent two NS natal kicks of $\sigma_{\rm NS}=100$ and $265\,{\rm km}\,{\rm s}^{-1}$. The top and right panels display {the probability density functions of} posteriors on the merger rate densities of {mgBHNS} and BHNS mergers obtained from \cite{Abbott2024Observation}. The $10\%$ and $50\%$ credible interval regions of the posteriors are marked as solid gray lines. The shadow region in the bottom right represents the prohibited parameter space where {$\mathcal{R}_{\rm mgBHNS}>\mathcal{R}_{\rm BHNS}$}. }
\label{fig:EventRate}
\end{figure} 

The simulated local event rate densities of {mgBHNS} and BHNS mergers are listed in Table \ref{tab:EventRate} and are displayed in Figure \ref{fig:EventRate}. {In Table \ref{tab:EventRate},} {we also calculate the confidence intervals for our simulated population synthesis rate densities to fall within} {the posterior distribution of $\mathcal{R}_{\rm mgBHNS}$ and $\mathcal{R}_{\rm BHNS}$ obtained from the population analyses reported by the LVK Collaboration \citep{Abbott2024Observation}} 
{, allowing us to evaluate which population synthesis models are better constrained based on the present observations. } The event rate densities of our population synthesis simulations are all within the {$90\%$} credible intervals of observed {$\mathcal{R}_{\rm mgBHNS}$} and $\mathcal{R}_{\rm BHNS}$. We find that the models with {a higher SN kick} indicate a better agreement with the LVK's observations. Among these population synthesis models, the model of {$\alpha2\sigma265$ exhibits the best consistency with the GW observations.}

\begin{figure}[t]
\centering
\includegraphics[width=1\linewidth, trim = 53 35 222 60, clip]{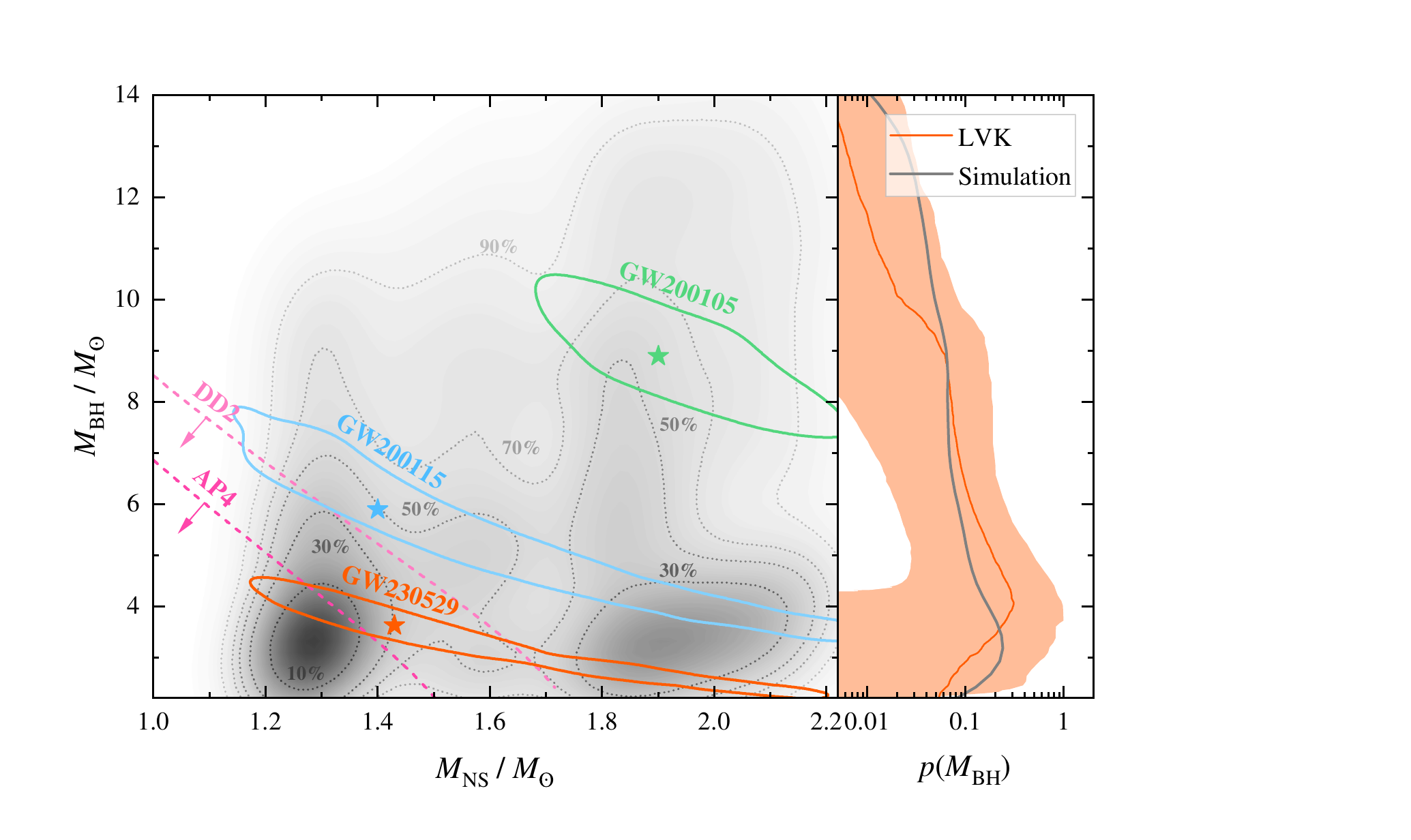}
\caption{NS and BH masses of our population synthesis simulations. The grey dotted lines represent the credible interval regions of the simulation results, spanning from $10\%$ to $90\%$, estimated using kernel density estimation. The {source-frame} medians (star points) along with their corresponding $90\%$ credible intervals (solid lines) are illustrated for the posterior distributions of GW200105 (green), GW200115 (blue), and GW230529 (orange). The right panel shows {the probability density functions of} the population distribution of the BH mass inferred by the GW observations \citep[orange histogram; ][]{Abbott2024Observation}, contrasted with that derived from our population synthesis simulations (grey histogram). For an EoS of AP4 (DD2), 
non-spinning BHNS mergers with component masses located at the bottom left parameter space of the pink (light pink) dashed line can allow tidal disruption. }
\label{fig:PopulationBHandNSMass}
\end{figure} 

We show the distribution of the BH mass and NS mass for {the $\alpha2\sigma265$} population model in Figure \ref{fig:PopulationBHandNSMass}. The {source-frame} medians with the $90\%$ credible intervals of the inferred posterior samples for GW200105, GW200115, and GW230529 are also displayed. One can find that the model predicts that the majority of BHNS mergers can have a BH mass of {$2.2-13.5\,M_\odot$}, and a NS mass of $1.25-2.2\,M_\odot$, with a concentration of $\sim1.3\,M_\odot$. The inferred properties of GW200105, GW200115, and GW230529 lie within the {$70\%$} credible interval of our simulated BHNS population. In particular, the medians with a large fraction of the posterior for the mgBHNS event, i.e., GW230529, recently reported by the LVK Collaboration significantly overlap with the highest probability region of the simulated BHNS population. We find that the predicted BH mass distribution is basically consistent with that of GW BHNS population inferred by the presently detected three high-confidence BHNS mergers. Compared with the current GW observations, where the BH mass peaks at $\sim4\,M_\odot$, our simulated BH population has a lower mass peak of $3.4\,M_\odot$, {which is approximately at the median of the posterior BH mass distribution of GW230529.}

Overall, the posterior masses of GW230529 given by the LVK Collaboration are close to the highest probability region of the simulated BHNS population. Our population synthesis simulation results can simultaneously match the inferred event rate densities of mgBHNS and total BHNS mergers, along with the GW population distribution of the BH mass in BHNS mergers. {Because GW230529 contributes significantly to the inferred mgBHNS rate densities,} one can expect that GW230529 can likely originate from the isolated binary evolution channel.

\section{EM Counterparts of GW230529 and Implications for Future BHNS EM Observations}

GW230529 was only observed by LIGO Livingston and, hence, had a sky localization with a 90\% credible area of $\sim25600\,{\rm deg}^2$ \citep{Abbott2024Observation}, which was too wide for follow-up observations. There were no kilonova and GRB candidates reported after this event. In this section, we will study the tidal disruption probability and EM signals of GW230529. Furthermore, with the discovery of the existence of mgBHNS mergers, our previous understanding of the detectability of BHNS EM signals may change. We will also briefly explore the future detectability of BHNS EM signals based on our population synthesis simulations.

\subsection{Tidal Disruption Probability}

\begin{figure*}[t]
\centering
\includegraphics[width=0.470\linewidth, trim = 50 36 177 25, clip]{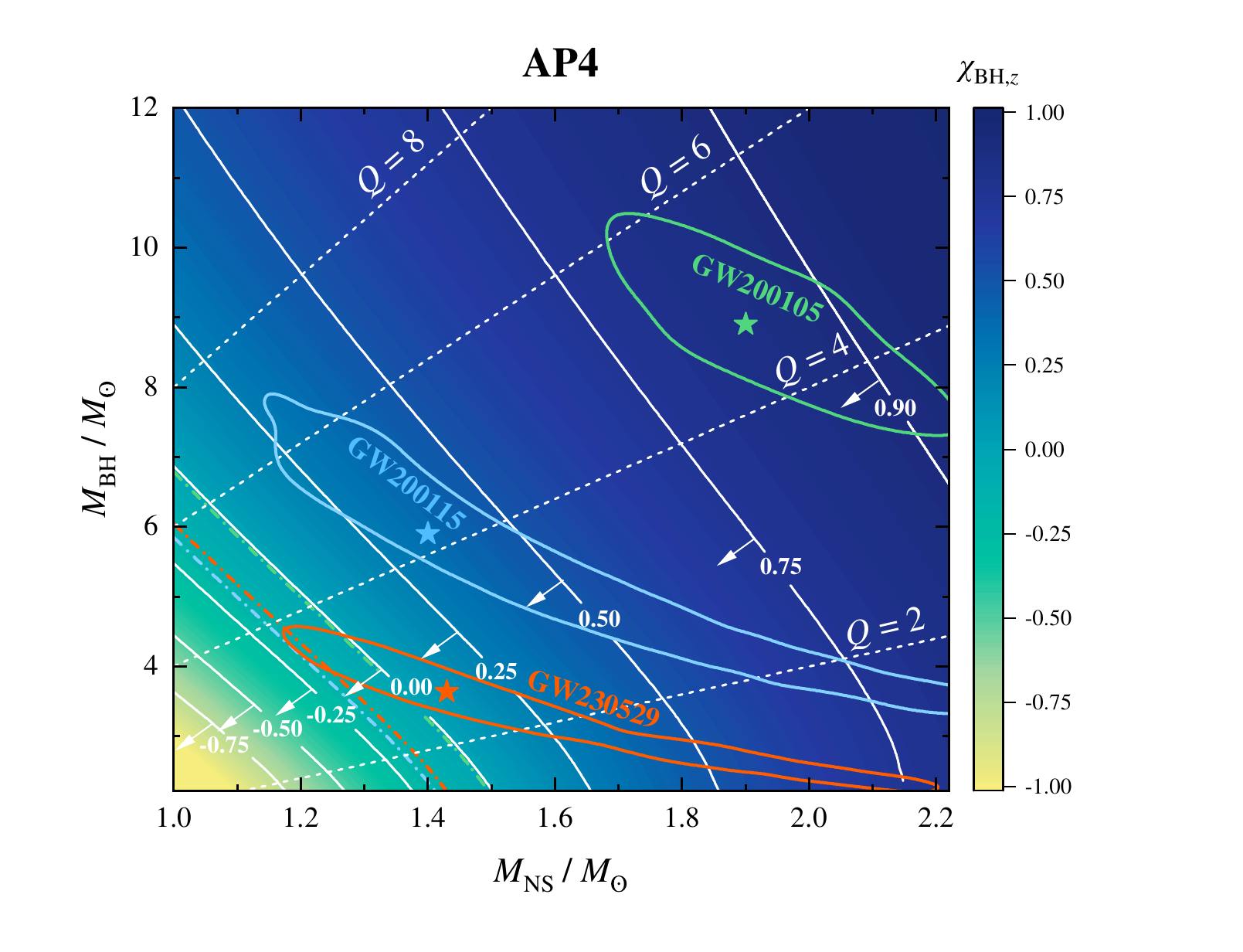}
\includegraphics[width=0.520\linewidth, trim = 75 36 94 25, clip]{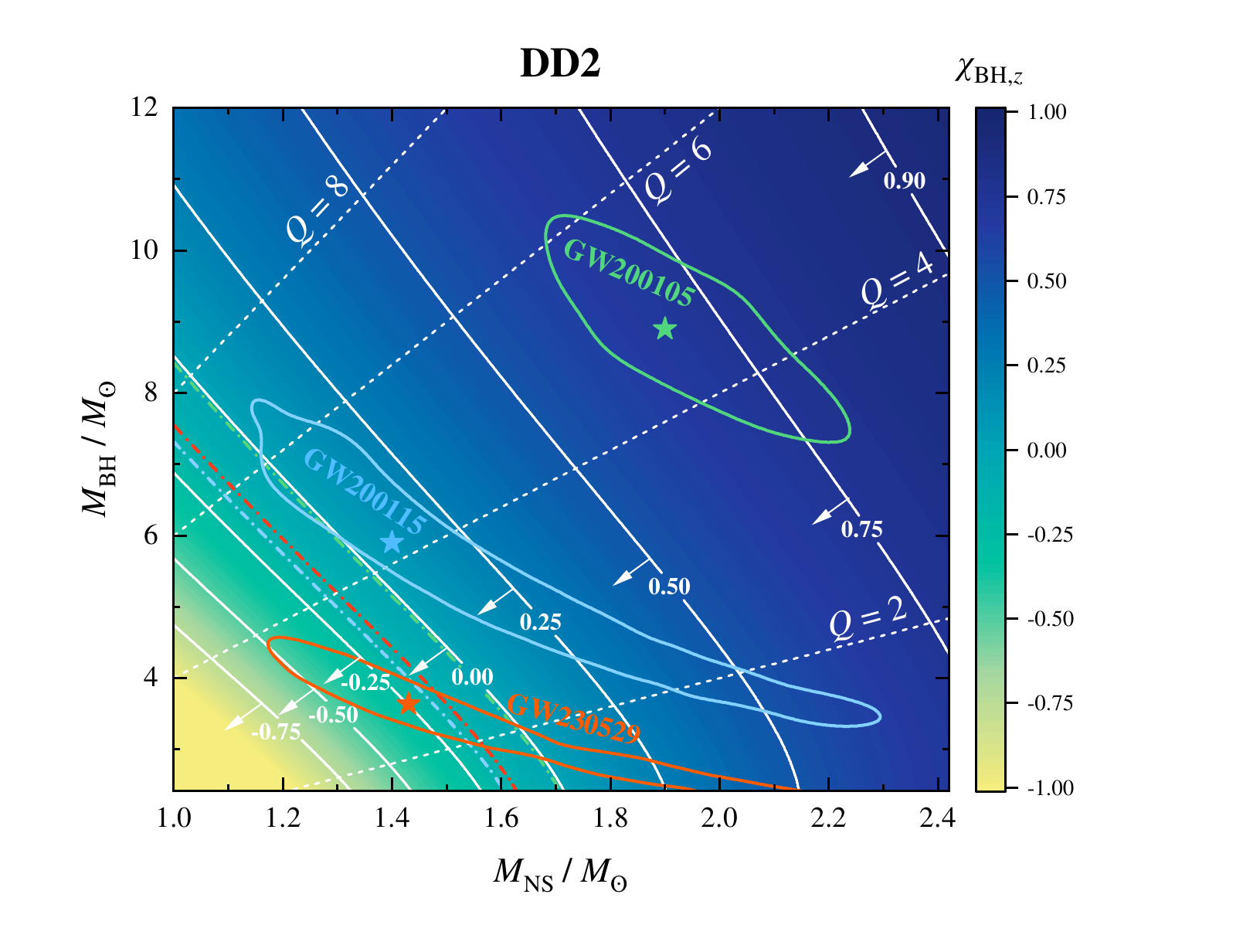}
\caption{{Source-frame mass parameter space} for BHNS merger systems to allow NS tidal disruption. We mark the mass ratio from $Q=2$ to 8 as dashed lines in each panel. The solid lines represent the primary BH aligned-spin $\chi_{{\rm BH},z}$ from $-0.75$ to $0.90$. For a specific $\chi_{{\rm BH},z}$, BHNS mergers with component masses located at the bottom left parameter space (denoted by the direction of the arrows) can undergo tidal disruption. For GW200105 (green), GW200115 (blue), and GW230529 (orange), the 90\% credible posterior distributions (colored solid lines) and the medians (colored stars) are displayed, {while the corresponding median values of $\chi_{{\rm BH},z}$ for these three sources are marked as dashed-dotted lines.}}
\label{fig:TDERegion}
\end{figure*}

Whether tidal disruption can occur in a BHNS merger can be described by an empirical formula \citep{Foucart2018}
\begin{equation}
\label{equ:TotalEjectaMassFunction}
    \frac{M_{\rm total,fit}}{M^{\rm b}_{\rm NS}} = \left[\max\left(\alpha\frac{1 - 2C_{\rm NS}}{\eta^{1 / 3}} - \beta \widetilde{R}_{\rm {ISCO}}\frac{C_{\rm NS}}{\eta} + \gamma , 0 \right)\right]^ {\delta},
\end{equation}
which is used to calculate the total remnant mass outside the remnant BH horizon, where $\alpha = 0.406$, $\beta = 0.139$, $\gamma = 0.255$, $\delta = 1.761$, $M_{\rm NS}^{\rm b}$ is the baryonic NS mass, $\eta = Q / (1 + Q) ^ 2$, $Q = M_{\rm BH} / M_{\rm NS}$ is the mass ratio between the BH mass $M_{\rm BH}$ and the NS mass $M_{\rm NS}$, {$C_{\rm NS} = GM_{\rm NS}/c^2R_{\rm NS}$ is the compactness dependent on the NS equation of state (EoS)} with the gravitational constant $G$, speed of light $c$, and NS radius $R_{\rm NS}$. The normalized inner stable circular orbit radius \citep{Bardeen1972} can be expressed as $\widetilde{R}_{\rm {ISCO}} = 3 + Z_2 - {\rm {sign}}(\chi_{{\rm BH},z})\sqrt{(3 - Z_1)(3 + Z_1 + 2Z_2)}$, where $Z_1 = 1 + (1 - \chi_{{\rm BH},z}^2) ^ {1 / 3} [(1 + \chi_{{\rm BH},z})^{1 / 3} + (1 - \chi_{{\rm BH},z})^{1 / 3}]$, $Z_2 = \sqrt{3 \chi_{{\rm BH},z}^2 + Z_1^2}$, and $\chi_{{\rm BH},z}$ is the dimensionless spin parameter projected onto the orientation of orbital angular momentum, abbreviated as the BH aligned-spin hereafter. This formula can be more accurately applied in the range of $Q \in [1 , 7]$, $\chi_{\rm BH} \in [-0.5 , 0.9]$, and $C_{\rm NS} \in [0.13 , 0.182]$ \citep{Foucart2018}. {BHNS mergers with component masses located in the parameter space of $M_{\rm total}>0$ can undergo tidal disruption and generate EM signals.}

{We consider two specific EoSs commonly used in the literature, among which AP4 \citep{Akmal1997} is one of the most probable EoSs with a Tolman-Oppenheimer-Volkoff mass of $M_{\rm TOV}=2.22\,M_\odot$, while DD2 \citep{Typel2010} is one of the stiffest EoSs constrained by GW170817 \citep{Abbott2018GW170817,Abbott2019Properties}, with $M_{\rm TOV}=2.42\,M_\odot$. We calculate the baryonic NS mass in Equation (\ref{equ:TotalEjectaMassFunction}) as follow: $M_{\rm NS}^{\rm b} = M_{\rm NS,\odot} + A_1\times{M}_{\rm NS,\odot}^2 + A_2{M}_{\rm NS,\odot}^3$, where $M_{\rm NS,\odot}=M_{\rm NS}/M_\odot$, and we adopt the fitting values $A_1 = 0.045$ ($0.046$) and $A_2 = 0.023$ ($0.014$) for an EoS of AP4 (DD2) from \cite{Gao2020}. The NS compactness is given by the fitting formula from \cite{Coughlin2017}, i.e., $C_{\rm NS} = 1.1056\times(M_{\rm NS}^{\rm b}/M_{\rm NS}-1)^{0.8277}$. }

Figure \ref{fig:TDERegion} shows the parameter space where the NS can be tidally disrupted using Equation (\ref{equ:TotalEjectaMassFunction}). For a given $\chi_{{\rm BH},z}$, BHNS mergers composed of a low-mass BH component and a low-mass NS component are more easily to have tidal disruption. The mass space that allows NS tidal disruption expands significantly with increasing $\chi_{{\rm BH},z}$ and the adoption of a stiffer EoS. We display the 90\% credible posterior distributions and the medians of component masses, as well as the medians of BH aligned-spins $\chi_{{\rm BH},z}$, for GW200105, GW200115, and GW230529 \citep{Abbott2021Observation,Abbott2024Observation} in Figure \ref{fig:TDERegion}. {Based on the observed GW samples reported by the LVK Collaboration \citep{Abbott2021Observation,Abbott2024Observation}, the explicit results of the tidal disruption probabilities for these three BHNS mergers are listed in Table \ref{tab:TDEandEM}.} It is obvious that GW200105 is unlikely to make tidal disruption to generate any EM counterparts, because its BH and NS components are massive with the mass distribution significantly outside the tidal disruption region. Although GW200115 has BH and NS masses much lighter than those of GW200105, tidal disruption can occur with only a very low probability of $P_{\rm tidal}=2.76\%$ under the EoS of DD2. Compared with GW200105 and GW200115 whose BH masses are $\gtrsim5\,M_\odot$, GW230529, as a mgBHNS merger, occupies a larger portion of its mass parameter space located within the disruption region, especially when considering a stiffer EoS. More specifically, the tidal disruption probabilities are $12.8\%$ and $63.2\%$ by adopting the EoS of AP4 and DD2, respectively. The former value is similar to that reported in \cite{Abbott2024Observation}, which was obtained by marginalizing over the EoS.

\begin{deluxetable*}{ccc|cc|cc}[tpb]
\tablecaption{Tidal Disruption Probability, Kilonova Brightness, and GRB Detection Probability of BHNS Mergers} \label{tab:TDEandEM}
\tablehead{
\multirow{2}{*}{GW Event} & 
\multirow{2}{*}{EoS} &
\multirow{2}{*}{$P_{\rm tidal}$} & 
\multirow{2}{*}{$m_{g}/{\rm mag}$} &
\multirow{2}{*}{$m_{r}/{\rm mag}$} & \multicolumn{2}{c}{$P_{\rm GRB}$} \\
&&&&&\colhead{$\theta_{\rm c}=3.5^\circ$} & \colhead{$\theta_{\rm c}=7^\circ$}
}
\startdata
\multirow{2}{*}{GW200105} & AP4 & 0\% & $-$ & $-$ & $-$ & $-$    \\
& DD2 & 0\% & $-$ & $-$ & $-$ & $-$ \\ \hline
\multirow{2}{*}{GW200115} & AP4 & 0\% & $-$ & $-$ & $-$ & $-$ \\
& DD2 & 2.76\% & $24.5^{+1.1}_{-1.3}$ & $24.4^{+1.3}_{-1.1}$ & $0.004\%$ & $0.004\%$\\ \hline
\multirow{2}{*}{GW230529} & AP4 & 12.8\% & $24.4^{+1.5}_{-1.9}$ & $24.3^{+1.7}_{-1.9}$  & $0.35\%$ & $0.77\%$ \\ 
& DD2 & 63.2\% & $23.4^{+1.6}_{-1.3}$ & $23.2^{+1.6}_{-1.4}$ & 1.56\% & $4.61\%$
\enddata
\tablecomments{The columns are (1) the GW event; (2) the selected EoS; (3) the tidal disruption probability; (4) the median values with $90\%$ credible intervals of $g$-band apparent magnitude distribution; (5) the median values with $90\%$ credible intervals of $r$-band apparent magnitude distribution; (6) the GRB detection probability by considering the jet core opening angles of $\theta_{\rm c}=3.5^\circ$ and $7^\circ$. }
\end{deluxetable*}

\subsection{Kilonova Properties}

\begin{figure*}[t]
\centering
\includegraphics[width=0.5155\linewidth, trim = 81 29 90 68, clip]{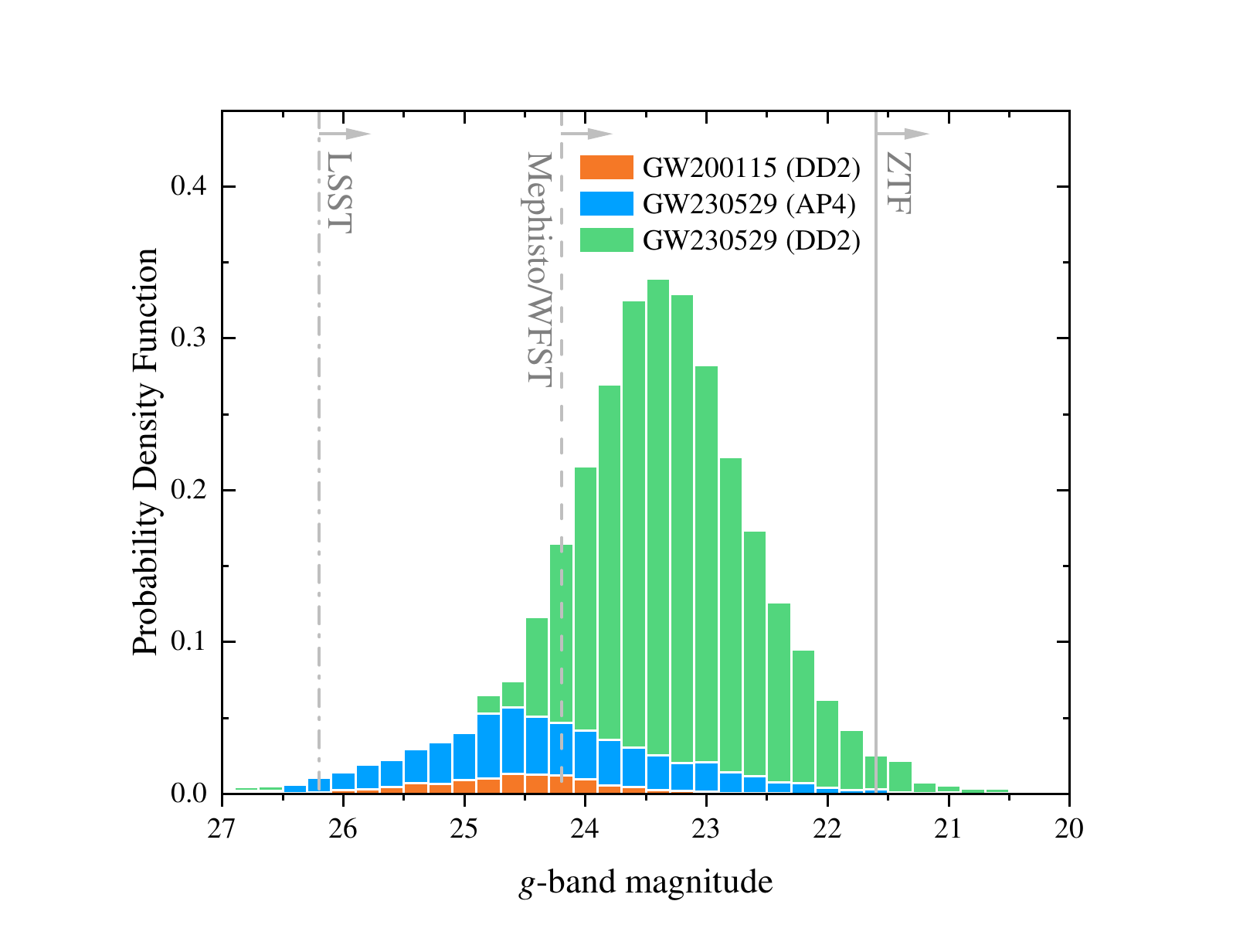}
\includegraphics[width=0.4745\linewidth, trim = 129 29 90 68, clip]{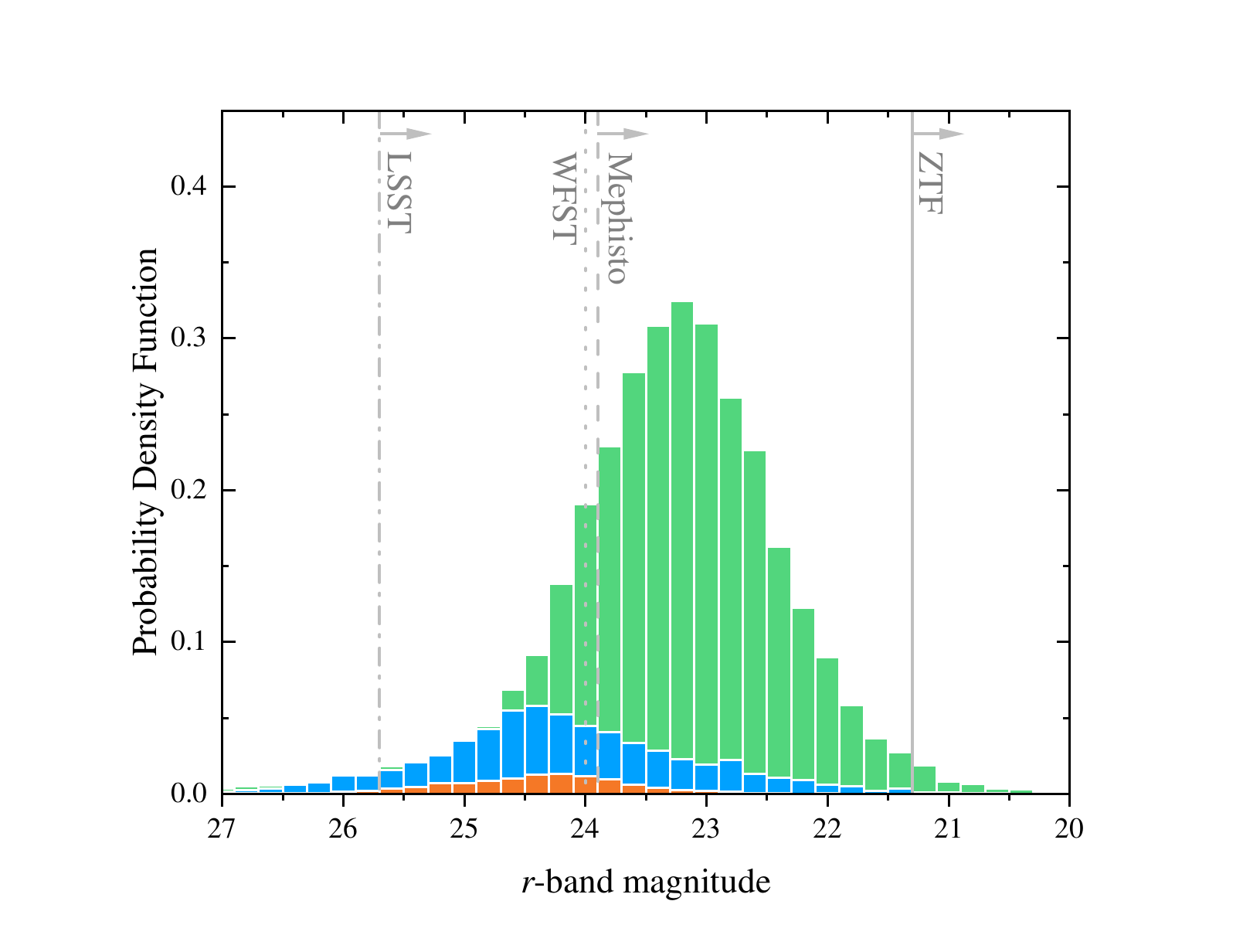}
\caption{Probability density distributions of BHNS merger kilonova $g$-band (left panel) and $r$-band (right panel) peak magnitude for GW200115 and GW230529. The orange histograms represent the probability density for GW200115 with the consideration of the AP4 EoS, while the blue and green histograms show the probability densities for GW230529, employing the AP4 and DD2 EoSs, respectively. The gray solid, dashed, dotted, and dashed-dotted lines show the threshold depths of ZTF, Mephisto, WFST, and LSST for an exposure time of $300\,{\rm s}$ under the ideal observing conditions \citep{Zhu2023}. Here, the bin width of the histograms is set as $\Delta = 0.2\,{\rm mag}$.}
\label{fig:LuminosityDistribution}
\end{figure*}

Tidal disruption of a BHNS merger can directly generate unbound lanthanide-rich dynamical ejecta, while other unswallowed material can form a disk around the remnant BH. Based on \cite{Zhu2020}, one can calculate the mass of the dynamical ejecta by $M_{\rm d}=\min(f_{\rm max}M_{\rm total,fit}, M_{\rm d,fit})$ and disk mass by $M_{\rm disk}=M_{\rm tot,fit}-M_{\rm d}$, where $f_{\rm max}$ represents the maximum fraction of the dynamical ejecta mass in the total remnant mass as determined by the numerical relativity simulations \citep{Kyutoku2015} and $M_{\rm d,fit}$ can be obtained by substituting the fitting parameters in Equation (\ref{equ:TotalEjectaMassFunction}) with $\alpha=0.273$, $\beta=0.035$, $\gamma=-0.153$ and $\delta=1.491$. The root-mean-square velocity of the dynamical ejecta can be described by $v_{\rm rms,d}=(-0.441Q^{-0.224}+0.549)c$, which is applied for $Q\in[1,7]$. In addition to the tidal dynamical ejecta, the remnant disk can also produce lanthanide-free neutrino-driven wind ejecta caused by neutrino heating and intermediate-opacity viscosity-driven wind ejecta due to viscous heating and angular momentum transport. According to some numerical simulation results \citep[e.g.,][]{Fernandez2015, Just2015,Siegel2017}, neutrino-driven ejecta and viscosity-driven ejecta can have masses of $\sim1\%$ and $\sim20\%$ of the disk mass, respectively, with the root-mean-square velocities of $\sim0.03\,c$ and $\sim0.667\,c$. We set the gray opacities of neutrino-driven ejecta, and viscosity-driven ejecta as $1\,{\rm cm}^2{\rm g}^{-1}$, $5\,{\rm cm}^2{\rm g}^{-1}$, and $20\,{\rm cm}^2{\rm g}^{-1}$ \citep{Tanaka2020}.

We use the detailed viewing-angle-dependent BHNS kilonova model presented by \cite{Zhu2020}, which is essentially consistent with other models and simulations in the literature \citep[e.g.,][]{Kawaguchi2016,Kawaguchi2020,Barbieri2019,Darbha2021,Zhu2022Long,Gompertz2023}, to simulate multi-band lightcurves of kilonova associated with BHNS GWs. The input parameters include binary parameters ($M_{\rm BH}$, $M_{\rm NS}$, and $\chi_{{\rm BH},z}$) which can be obtained from the GW posterior results, EoS (i.e., AP4 and DD2), luminosity distance $D_{\rm L}$, redshift $z$, and two viewing angle parameters (including the latitudinal viewing angle $\theta_{\rm view}$ and the longitudinal viewing angle $\varphi_{\rm view}$). Here $\theta_{\rm view}$ equals the zenith angle between the total angular momentum and the line of sight $\theta_{JN}$ inferred from the observations of GW230529, while $\varphi_{\rm view}$ is randomly simulated between 0 and $2\pi$.

In Figure \ref{fig:LuminosityDistribution}, we show the probability density distributions of $g$- and $r$-band peak apparent magnitudes for the GW230529 kilonova. The medians with $90\%$ credible intervals of the peak magnitude distributions are listed in Table \ref{tab:TDEandEM}. We also mark the threshold depths of four representative survey projects, including the Zwicky Transient Facility \citep[ZTF;][]{Bellm2019}, the Multi-channel Photometric Survey Telescope (Mephisto\footnote{\url{http://www.mephisto.ynu.edu.cn}}), the Wide Field Survey Telescope \citep[WFST;][]{Wang2023} and the Large Synoptic Survey Telescope \citep[LSST;][]{LSST2009}, for an exposure time of $300\,{\rm s}$ in Figure \ref{fig:LuminosityDistribution}. The peak apparent magnitudes of the GW230529 kilonova exhibit broad distributions with the medians of $\sim24.3\,{\rm mag}$ and $\sim23.3\,{\rm mag}$ when adopting the EoSs of AP4 and DD2, respectively. The medians of apparent magnitude distribution for DD2 are $1\,{\rm mag}$ brighter than those of GW200115, because GW230529 with $D_{\rm L} = 197^{+107}_{-96}{\rm Mpc}$ is much closer than GW200115 with $D_{\rm L} = 310^{+150}_{-110}{\rm Mpc}$. Even though when considering a stiff EoS of DD2, we can find that ZTF could have a limited probability of observing the kilonova emission associated with GW230529. However, operational survey projects, including Mephisto and WFST, as well as future LSST can have enough detectability to discover the GW230529 kilonova if GW230529 can make tidal disruption.

\subsection{GRB Properties}

\begin{figure*}[t]
\centering
\includegraphics[width=0.520\linewidth, trim = 70 115 90 40, clip]{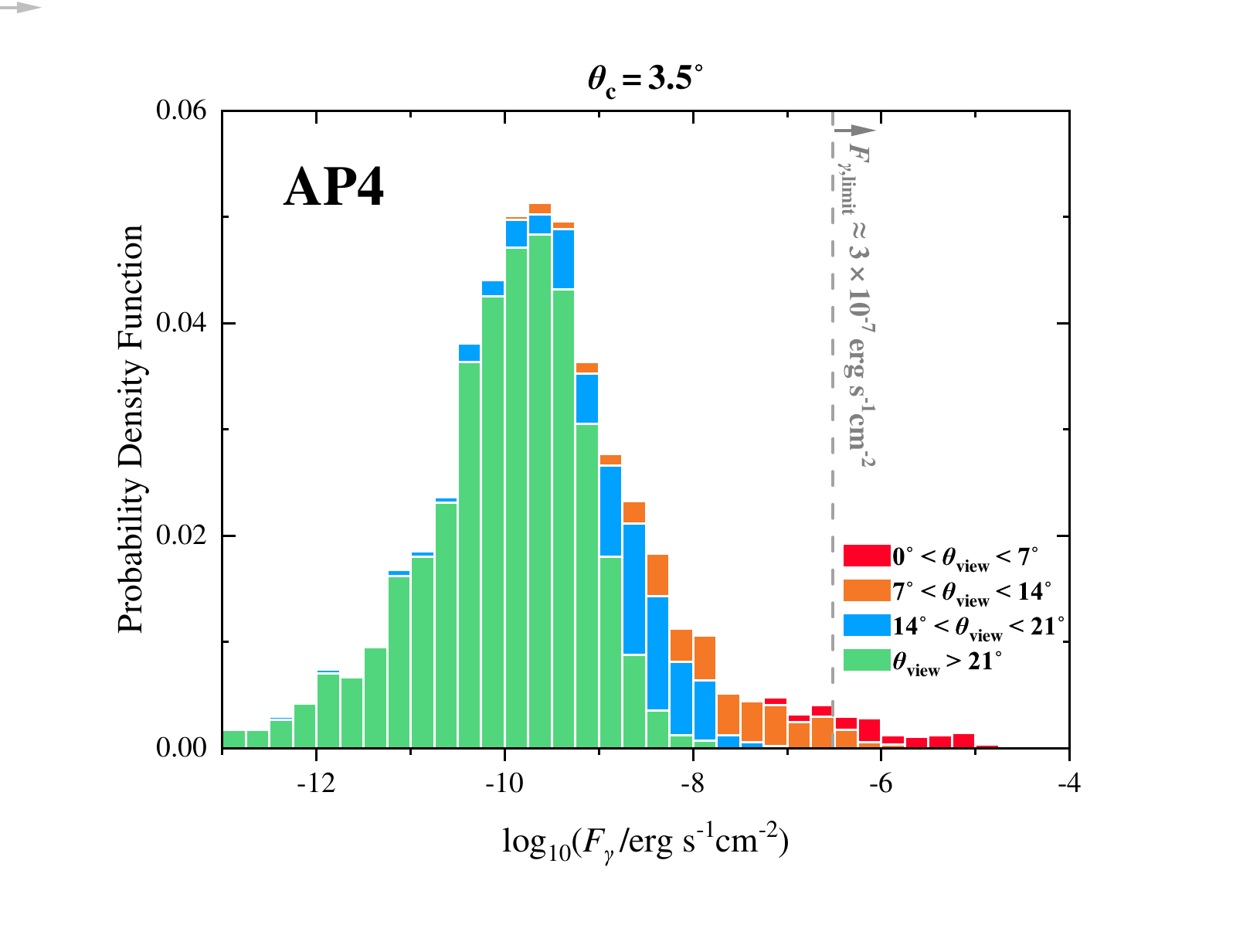}
\includegraphics[width=0.470\linewidth, trim = 129 115 90 40, clip]{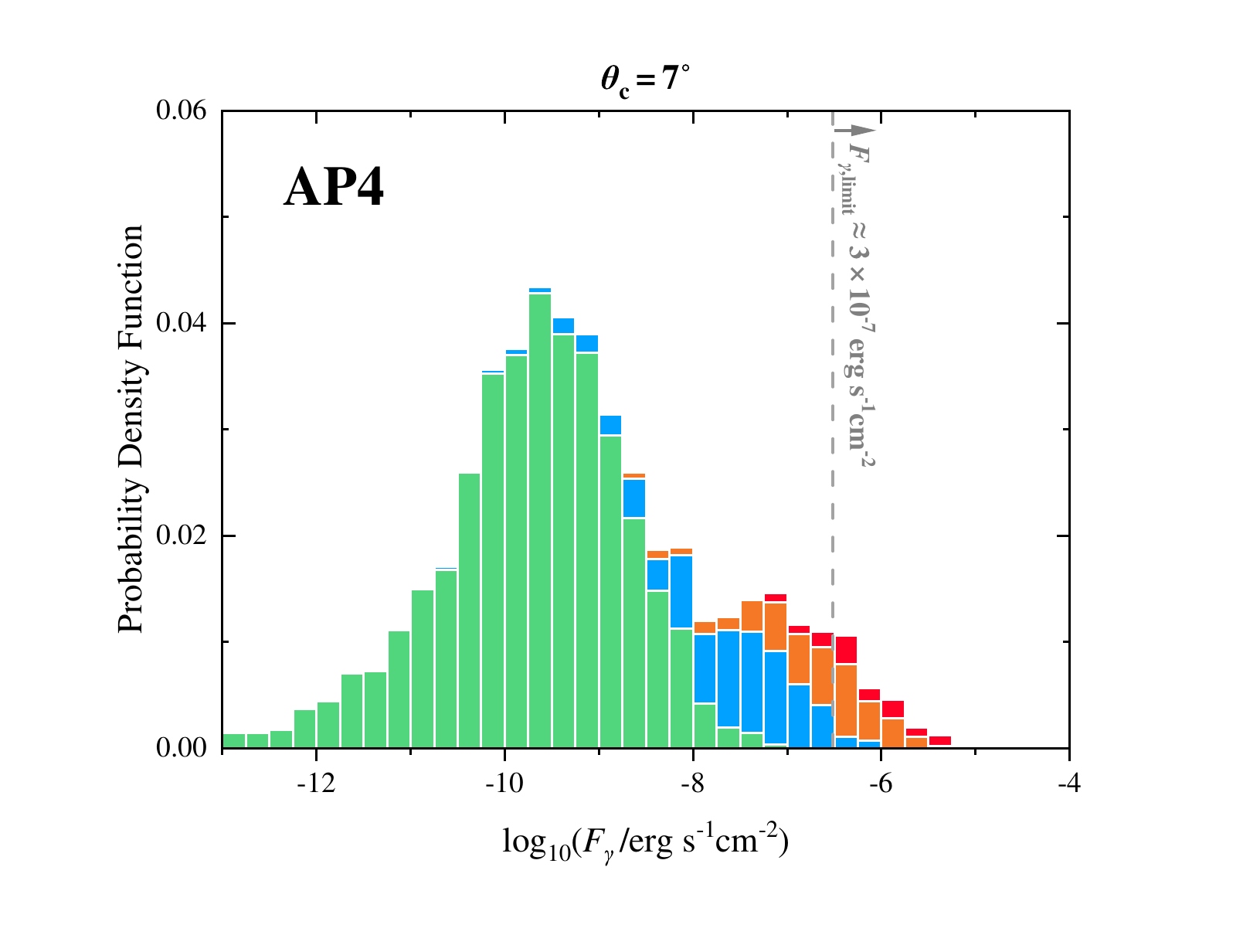}
\includegraphics[width=0.520\linewidth, trim = 70 50 90 60, clip]{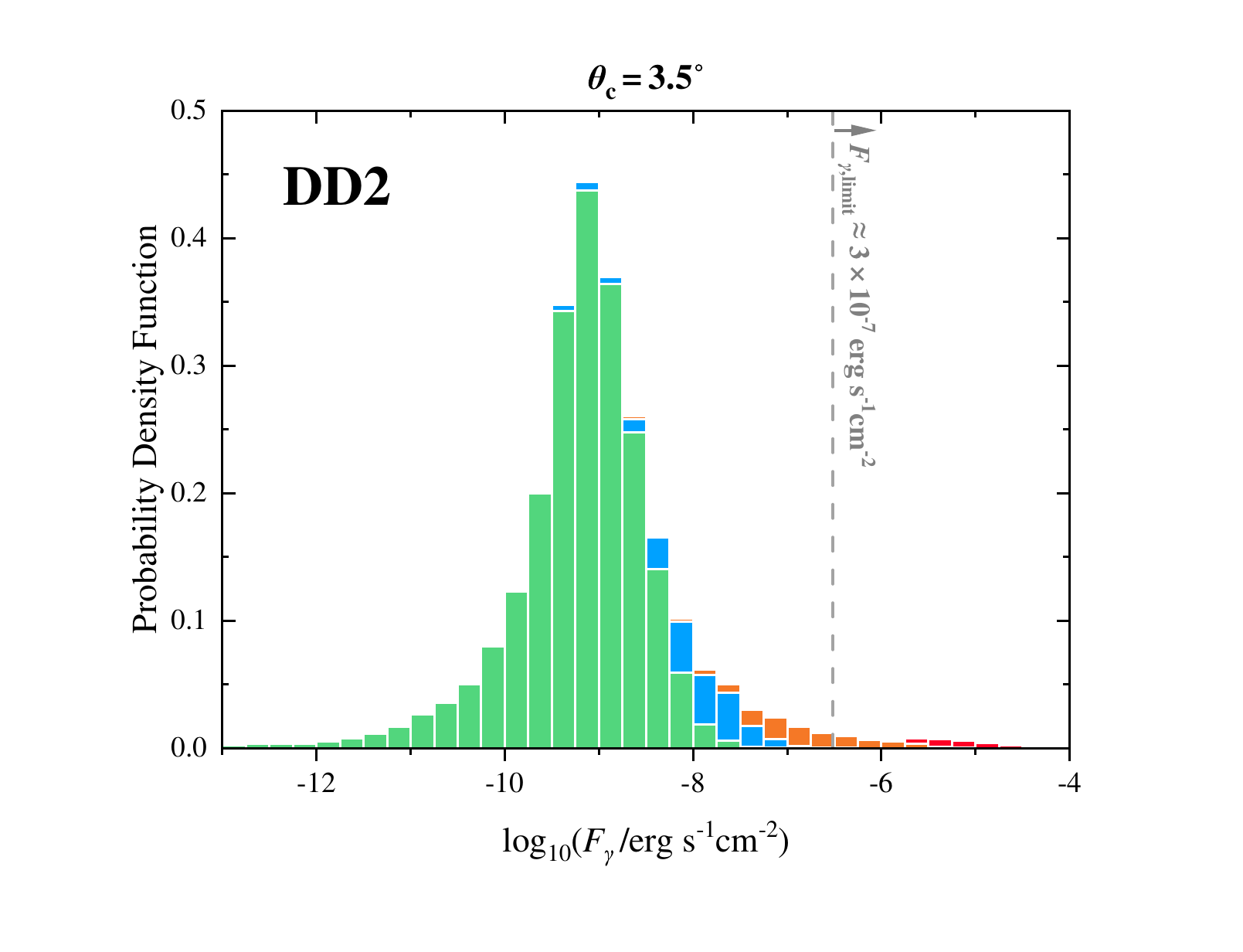}
\includegraphics[width=0.470\linewidth, trim = 129 50 90 60, clip]{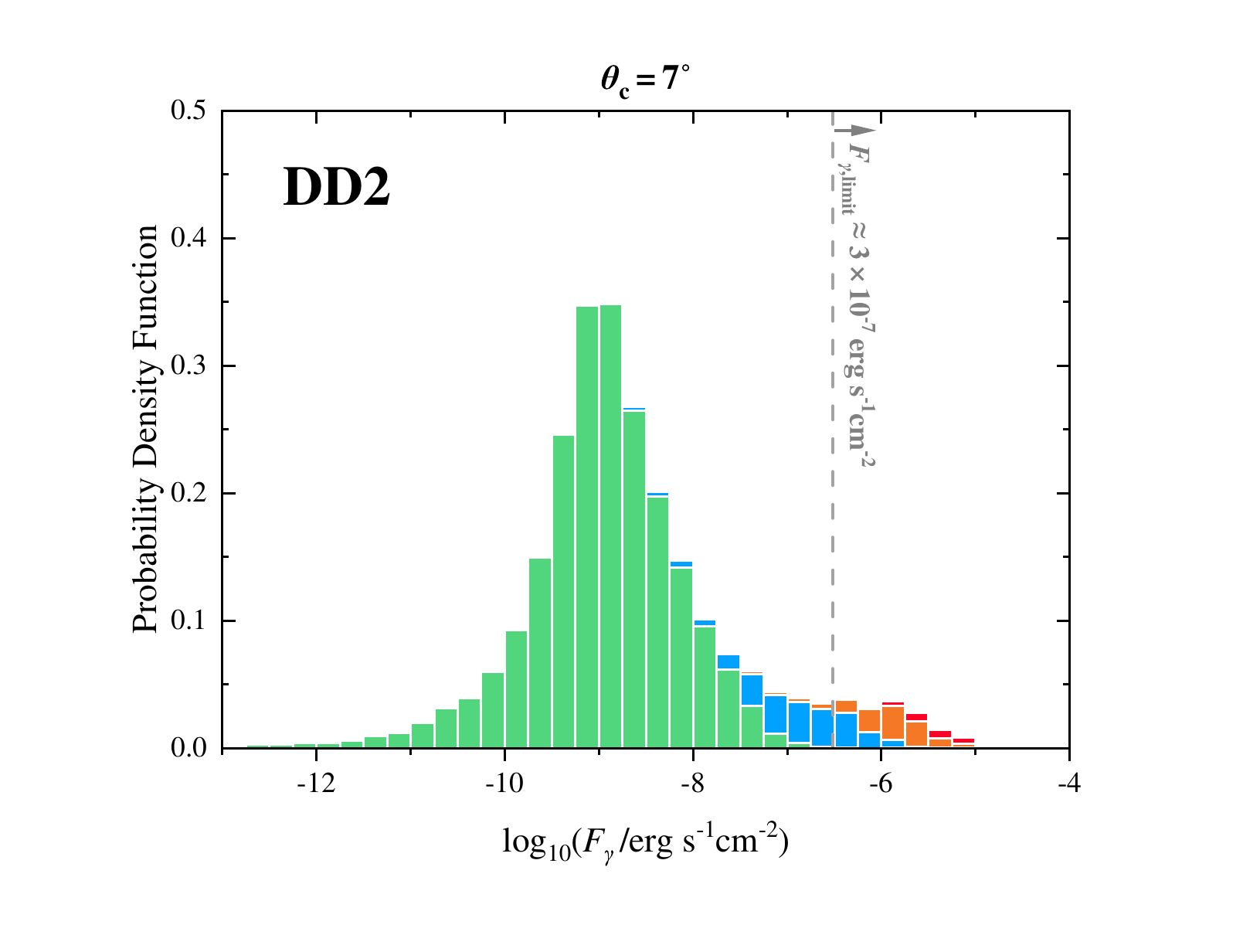}
\caption{Probability density distributions of BHNS merger GRB $\gamma$-ray flux for GW230529. Two different core opening angles are considered: $\theta_{\rm c}= 3.5^\circ$ (left panels) and $7^\circ$, as well as two EoSs, namely AP4 (top panels) and DD2 (bottom panels). The red, orange, blue, and green histograms represent GRB with latitudinal viewing angle ranges of $\theta_{\rm view}<7^\circ$, $7^\circ<\theta_{\rm view}<14^\circ$, $14^\circ<\theta_{\rm view}<21^\circ$, and $>21^\circ$, respectively. The gray dashed lines indicate the effective sensitivity limit for $\gamma$-ray detectors. We note that the $y$-axis scalings differ between the top and bottom panels.}
\label{fig:GRB}
\end{figure*}

We adopt a single-Gaussian structured jet model \citep{Zhang2002} to describe the GRB angular distributions of the jet kinetic energy and Lorentz factor, i.e.,
\begin{equation}
\begin{split}
    \frac{dE}{d\Omega}(\theta) &= \varepsilon_{\rm c}\exp\left(-\frac{\theta^2}{2\theta_{\rm c}^2}\right), \\
    \Gamma(\theta) &= \Gamma_{\rm c}\exp\left(-\frac{\theta^2}{2\theta_{\rm c}^2}\right) + 1,
\end{split}
\end{equation}
where $\Gamma_{\rm c}$ is the jet core Lorentz factor, $\theta_{\rm c}$ is the jet core opening angle, $\varepsilon_{\rm c} \approx E_{\rm K,jet} /2\pi\theta_{\rm c}^2$, and the kinetic energy of the Blandford–Znajek jet \citep{Blandford1977} is $E_{\rm K,jet} = \epsilon(1-\xi)M_{\rm disk}c^2\Omega_{\rm H}^2f(\Omega_{\rm H})$ with $\epsilon\approx0.015$ following \cite{Barbieri2019}, the disk mass loss ratio $\xi\approx0.2$ \citep{Fernandez2015,Just2015,Siegel2017}, the dimensionless angular frequency at the BH horizon $\Omega_{\rm H}=\chi_{\rm BH,f}/2\left({1+\sqrt{1-\chi^2_{\rm BH,f}}}\right)$, the dimensionless spin of the final BH $\chi_{\rm BH,f}$ calculated by using Equation (11) from \cite{Pannarale2013}, and the high-spin correction $f(\Omega_{\rm H})=1+1.38\Omega_{\rm H}^2-9.2\Omega_{\rm H}^4$. The isotropic equivalent energy of the GRB prompt emission is given by \citep{Salafia2015}
\begin{equation}
    E_{\gamma,{\rm iso}} = \eta_\gamma\int\frac{dE/d\Omega}{\Gamma^4(1-\beta\cos\alpha)^3}d\Omega(\theta,\varphi),
\end{equation}
where the radiation efficiency is typically adopted to $\eta_\gamma = 0.1$, $\beta = (1-\Gamma^{-2})^{1/2}$, $\cos\alpha = \cos\theta\cos\theta_{\rm view}+\sin\theta\sin\theta_{\rm view}\cos\varphi$. Therefore, we can obtain the $\gamma$-ray flux by $F_\gamma = E_{\rm \gamma,{\rm iso}}/4\pi D_{\rm L}^2t_{\rm j}$, where the jet duration is set to $t_{\rm j}\approx1{\rm s}$. Here, the GRB jet from BHNS mergers is assumed to be always launched towards the orbital angular momentum direction. 

Based on the observations of both prompt and afterglow emissions from GW170817, we assume that the jet core Lorentz factor and opening angle are $\Gamma_{\rm c}\sim500$ and $\theta_{\rm c}\sim3.5^\circ$ \citep[e.g.,][]{Lyman2018,Troja2020,Cao2023}, respectively. Furthermore, because the amount of ejecta along the polar direction of BHNS mergers is typically much less than that of BNS mergers, the jets from BHNS mergers might not be effectively collimated. Therefore, we explore the variation of the jet core opening angle of $\theta_{\rm c}=7^\circ$, assessing its influence on the detection of GRBs associated with GW230529.

The viewing angle of GW230529 inferred by the LVK Collaboration is $\theta_{\rm view} = 39^{+33}_{-26}$, suggesting that the associated GRB could be more likely an off-axis event. Based on this inferred viewing angle distribution, we model the $\gamma$-ray flux distributions of GRB associated with GW230529 by considering two EoSs in Figure \ref{fig:GRB}. The GRBs with flux larger than the effective sensitivity limit in $1-10^4\,{\rm keV}$ for Swift-BAT \citep{Gehrels2004} and SVOM-ECLAIRS \citep{Gotz2014}, i.e., $F_{\gamma,\rm limit}\sim3\times10^{-7}{\rm erg}\,{\rm s}^{-1}{\rm cm}^{-2}$ \citep{Song2019}, are assumed to be triggered by GRB detectors. The detection probabilities of associated GRBs for GW200115 and GW230529 are listed in Table \ref{tab:TDEandEM}.

As shown in Figure \ref{fig:GRB}, the flux distributions of GW230529-associated GRB are mostly much lower than the $\gamma$-ray sensitivity limit, peaking at $\sim10^{-10}-10^{-9}\,{\rm erg}\,{\rm s}^{-1}{\rm cm}^{-2}$. If $\theta_{\rm c}=3.5^\circ$, GRB detection typically requires viewing angles of $\theta_{\rm view}<7-14^\circ$; if $\theta_{\rm c}=7^\circ$, the allowed viewing angles can be slightly larger with $\theta_{\rm view}<14-21^\circ$. When adopting an EoS of AP4, the probabilities of GRB detection could be always lower than $1\%$. By considering the stiffer EoS of DD2, more BHNS mergers can have tidal disruption, leading to a more massive disk and hence brighter GRBs. Nevertheless, since the GW observation suggested that GW230529 was likely to be an off-axis event, the detection of its associated GRB remains challenging. When adopting  $\theta_{\rm c}=3.5^\circ$, the probability of GRB detection could be as low as $1.35\%$. Increasing $\theta_{\rm c}$ to $7^\circ$ can raise the detection probability to $4.61\%$. If GW230529 is a disrupted event, one may conclude that the absence of a detected GRB associated with GW230529 is likely due to it being an off-axis event, as indicated by the GW observation.

\subsection{Implication of EM detectability for BHNS Mergers}

\begin{deluxetable*}{cc|cc|ccc|cc}[tpb]
\tablecaption{{EM Detectability of BHNS Mergers in $300\,{\rm Mpc}$}} \label{tab:EM}
\tablehead{ 
\multirow{2}{*}{EoS} &
\multirow{2}{*}{Population} &
\multirow{2}{*}{$P_{\rm tidal}$} & 
\multirow{2}{*}{$R_{\rm tidal}/{\rm yr}^{-1}$} &
\multicolumn{3}{c}{$R_{\rm kilonova}/{\rm yr}^{-1}$} & \multicolumn{2}{c}{$R_{\rm GRB}/{\rm yr}^{-1}$} \\
&&&& \colhead{$22\,{\rm mag}$} & \colhead{$24\,{\rm mag}$} & \colhead{$26\,{\rm mag}$} & \colhead{$\theta_{\rm c}=3.5^\circ$} & \colhead{$\theta_{\rm c}=7^\circ$}
}
\startdata
\multirow{3}{*}{AP4} & Mass-gap BHNS & $17.1\%$ & $0.71$ & $0.01$ & $0.15$ & $0.70$ & $0.12$ & $0.18$\\
& High-mass BHNS & $0$ & $0$ & $0$ & $0$ & $0$ & $0$ & $0$ \\
& Total BHNS & $17.1\%$ & $0.71$ & $0.01$ & $0.15$ & $0.70$ & $0.12$ & $0.18$\\\hline
\multirow{3}{*}{DD2} & Mass-gap BHNS & $24.9\%$ & $1.04$ & $0.06$ & $0.79$ & $1.03$ & $0.20$ & $0.33$ \\
& High-mass BHNS & $3.4\%$ & $0.14$ & $0.01$ & $0.05$ & $0.14$ & $0.02$ & $0.02$ \\
& Total BHNS & $28.3\%$ & $1.18$ & $0.07$ & $0.84$ & $1.17$ & $0.22$ & $0.35$
\enddata
\tablecomments{The columns are (1) the selected EoS; (2) the BHNS population, including mgBHNS mergers, high-mass BHNS mergers with BH mass of $\gtrsim5\,M_\odot$, and total BHNS mergers (3) the tidal disruption probability; (4) the tidal disruption rate in $300\,{\rm Mpc}$; (5) the $g$-band kilonova detectable rate of BHNS mergers in $300\,{\rm Mpc}$ for three different detection depths of $m_g=22,\,24,\,{\rm and}\,26\,{\rm mag}$; (6) the detection rate of GRBs from BHNS mergers in $300\,{\rm Mpc}$ by considering two jet core opening angles of $\theta_{\rm c}=3.5^\circ$ and $7^\circ$. }
\end{deluxetable*}

In the standard binary evolution scenario, most BHs typically formed at wide orbits are expected to usually possess low spins \citep{Qin2018,Fuller2019,Belczynski2020,Mandel2021GW200115}. BHs with high orbital aligned spin in BHNS mergers can originate from tidal-induced spin-up \citep[e.g.,][]{Qin2018,Bavera2020,Hu2022,Chattopadhyay2022} or accretion-induced spin-up \citep[e.g.,][]{Wang2024,Xing2024,Zhu2024}, which may only account for a small fraction of BHNS mergers. For simplicity, we assume that the spins of BHs in our simulated BHNS mergers are near-zero. Furthermore, we randomly generate $\sin\theta_{\rm view}$ and $\varphi_{\rm view}$ within the ranges of $[0,1]$ and $[0,2\pi]$, respectively, for these simulated events. In Figure \ref{fig:PopulationBHandNSMass}, we show the mass distribution of our population synthesis simulations introduced in Section \ref{sec:Results}, as well as the tidal disruption region for BHNS mergers with non-spinning BH components. One can see that BHNS mergers located in the highest probability region are always allowed to have tidal disruption. When considering the NS EoS of AP4, all BHNS mergers with BH mass higher than the mass gap are solely plunging events, whereas tidal disruption can only happen for mergers between an mgBH and a $\lesssim1.4\,M_\odot$ NS. For the stiffer EoS of DD2, a fraction of NSs with masses $\lesssim1.4\,M_\odot$ can be tidally disrupted by $5-8\,M_\odot$ BHs. However, mgBHNS mergers would still be the dominant events causing disruption.

Based on our $\alpha2\sigma265$ population synthesis simulation result, we then simulate the tidal disruption probability, kilonova detectable rate, and GRB detection rate for GW BHNS mergers detected in the near future GW observing era (e.g., O4b and O5). We conservatively consider BHNS mergers occurring within $300\,{\rm Mpc}$, which is the observed distance of GW200105 and GW200115, as well as the representative distance of detectable BHNS mergers in O4 \citep{Abbott2018Prospects,Zhu2021Kilonova,Colombo2023,Gupta2023}. {The results are listed in Table \ref{tab:EM} without considering the simulated errors.} We find that $100\%$ and $\gtrsim90\%$ disrupted BHNS mergers are expected to originate from mgBHNS mergers by adopting the EoS of AP4 and DD2, respectively. When the detection depth exceeds $24\,{\rm mag}$, the survey projects can cover the majority of kilonovae associated with BHNS mergers. However, considering the rapid rise and decline of kilonovae, their signals may be challenging to be identified. Additionally, we find that the detection rate of GRBs originating from mgBHNS mergers is {$\sim0.1-0.4\,{\rm yr}^{-1}$}. Therefore, in the future, it may be possible to search for kilonova signals via the target-of-opportunity follow-up observations of both GW and GRB triggers, like the multimessenger observations between GW170817/GRB170817A/AT2017gfo.

\section{Conclusion}

In this paper, we use  \texttt{COMPAS} to explore the formation of GW230529, which is a mgBHNS merger between a $3.6^{+0.7}_{-1.2}M_\odot$ BH and a $1.43^{+0.59}_{-0.19}M_\odot$ NS reported by the LVK Collaboration. By adopting the `delayed' SN prescription, our population synthesis simulations can simultaneously match the inferred event rate densities of the mgBHNS and total BHNS mergers {obtained from the population analyses}, along with the population distribution of the BH mass in BHNS mergers modeled by the LVK Collaboration. The mass posterior of GW230529 significantly overlaps with the highest probability region of the simulated BHNS population. {Furthermore, mgBHNS mergers can also originate from dynamical formation in dense stellar environments or triple system, but the expected rate densities could be much lower than inferred GW mgBHNS rate densities \citep[e.g.,][]{Fragione2020,Gupta2020,Ye2020,Tagawa2021}. Since GW230529 contributes significantly to the current inferred mgBHNS rate densities}, we suggest that GW230529 could originate from the isolated binary evolution channel. 

By considering two specific EoSs of AP4 and DD2, the probabilities that GW230529 can have tidal disruption are $12.8\%$ and $63.2\%$, respectively. If GW230529 is a disrupted event, the associated kilonova is predicted to have an apparent magnitude of $\sim23-24\,{\rm{mag}}$, and hence, can be detected by the present survey projects and the LSST in the future. Since GW230529 could be an off-axis event inferred from the GW observation, its associated GRB might be too dim to be observed by $\gamma$-ray detectors, interpreting the lack of a detected GRB associated with the event. {Our results are generally consistent with those suggested by other recent studies \citep[e.g.,][]{Chandra2024,Ronchini2024}.}

The discovery of GW230529 revealed that the mass gap between $\sim2.2-5\,M_\odot$ may not exist. Previous studies indicated that most cosmological BHNS mergers involving BH masses of $\gtrsim5\,M_\odot$ are usually hard to make tidal disruption and to generate bright EM signals \citep[e.g.,][]{Zhu2021No,Zhu2021Kilonova,Zhu2022Population,Fragione2021,Drozda2022}. The existence of mgBHNS mergers suggests that BHNS mergers are still likely to be multimessenger sources that emit GWs, GRBs, and kilonovae. Although mgBHNS mergers account for {$\sim50\%$} of the cosmological BHNS population, we find that $100\%$ and $\gtrsim90\%$ disrupted BHNS mergers are expected to originate from mgBHNS mergers, when considering the EoSs of AP4 and DD2, respectively.

\software{\texttt{COMPAS} {\citep[version 02.39.00;][]{TeamCOMPAS2022}}; {\texttt{GWOSC}, https://gwosc.org;} \texttt{Python}, \url{https://www.python.org}; \texttt{Matlab}, \url{https://www.mathworks.com}; \texttt{OriginPro}, \url{https://www.originlab.com/originpro}}

\acknowledgments {Population Synthesis Simulations in this paper made use of the COMPAS rapid binary population synthesis code (version 02.39.00), which is freely available at \url{http://github.com/TeamCOMPAS/COMPAS}. We thank the GWOSC (Gravitational Wave Open Science Center; \url{https://gwosc.org}) for making the data of GW230529 publicly available at \url{https://doi.org/10.7935/6k89-7q62}. We thank Floor Broekgaarden for publicly providing access to a template of a table that can summarize the binary population synthesis settings (\url{https://github.com/FloorBroekgaarden/templateForTableBPSsettings}).} The authors acknowledge anonymous referees for useful discussions.  We thank {Floor Broekgaarden}, Paul D. Lasky, Liang-Duan Liu, Ilya Mandel, Eric Thrane, Xiangyu Ivy Wang, {Zepei Xing}, and Yun-Wei Yu for helpful discussions. JPZ thanks the COMPAS group at Monash University. JPZ acknowledges support from the Australian Research Council Centre of Excellence for Gravitational Wave Discovery (OzGrav), through project number CE17010004. YK and LS were supported by the Beijing Natural Science Foundation (1242018), the National SKA Program of China (2020SKA0120300), and the Max Planck Partner Group Program funded by the Max Planck Society. YQ acknowledges the support of Anhui Provincial Natural
Science Foundation (grant No. 2308085MA29).

\appendix
{We summarize the binary population synthesis settings of our \texttt{COMPAS} simulations in Table \ref{tab:COMPAS}.}

\begin{table*}
\caption{{Initial values and default settings of the population synthesis  simulation with {\texttt{COMPAS}}  }}
\label{tab:COMPAS}
\centering
\resizebox{\textwidth}{!}{\begin{tabular}{lll}
\hline  \hline
Description and name & Value/range & Note/setting   \\ 
\hline  \hline
\multicolumn{3}{c}{Initial conditions} \\ 
\hline
Initial primary mass $M_{1,\rm i}$ & $[5, 150]\,M_\odot$  & \citet{Kroupa2001} IMF  $\propto  {M_{1,\rm i}}^{-\alpha_{\rm IMF}}$  with $\alpha_{\rm{IMF}} = 2.3$ for stars above $5\,M_\odot$	  \\
Initial mass ratio $q_{\rm i} = M_{2,\rm i} /  M_{1,\rm i} $ & $[0, 1]$ & We assume a flat mass ratio distribution  $p(q_{\rm i}) \propto  1$ with $M_{2,\rm i}\geq 0.1\,M_\odot$  \\
Initial semi-major axis $a_{\rm i}$ & $[0.01, 1000]\,{\rm AU}$ & Distributed flat-in-$\log p(a_{\rm i}) \propto 1 / {a_{\rm i}}$ \\   
Initial metallicity $Z_{\rm i}$ & $[0.0001, 0.03]$ & Distributed using a uniform grid in $\log (Z_{\rm i})$ with 10 metallicities        \\
Initial orbital eccentricity $e_{\rm i}$ & 0 & All binaries are assumed to be circular at birth  \\
\hline
\multicolumn{3}{c}{Fiducial parameter settings:} \\ 
\hline
Stellar winds  for hydrogen rich stars & \cite{Belczynski2010} & Based on \cite{Vink2005}, including LBV wind mass loss with $f_{\rm{LBV}} = 1.5$   \\
Stellar winds for helium stars &  \cite{Belczynski2010b} & Based on   \cite{Hamann1998} and  \cite{Vink2005}  \\
Max transfer stability criteria & $\zeta$-prescription & Based on \cite{Vigna2018} and references therein     \\ 
Mass transfer accretion rate & thermal timescale & Limited by thermal timescale for stars  \cite{Vink2005,Vinciguerra2020} \\ 
 & Eddington-limited  & Accretion rate is Eddington-limit for compact objects  \\
Non-conservative mass loss & isotropic re-emission &  {\citet[][]{Massevitch1975,Bhattacharya1991}} \\ 
& &  {\cite{Soberman1997,Tauris2006}} \\
Case BB mass transfer stability & always stable & Based on  \citet{Tauris2015,Tauris2017,Vigna2018} \\ 
CE prescription & $\alpha_{\rm CE}-\lambda$ & Based on  \citet{Webbink1984,DeKool1990}  \\
$\alpha_{\rm CE}$-parameter & 1, 2, 5, 10 &  \\
CE $\lambda$-parameter & $\lambda$ & Based on \cite{Xu2010} and \cite{Dominik2012}  \\
Hertzsprung gap (HG) donor in {CE} & pessimistic & Defined in \citet{Dominik2012}: HG donors don't survive a {CE} phase \\
{SN} natal kick magnitude $v_{\rm k}$ & $[0, \infty)\,{\rm km}\,{\rm s}^{-1}$ & Drawn from Maxwellian distribution with standard deviation $\sigma_{\rm{rms}}$          \\
{SN} natal kick polar angle $\theta_{\rm k}$ & $[0, \pi]$ & $p(\theta_{\rm k}) = \sin(\theta_{\rm k})/2$ \\
{SN} natal kick azimuthal angle $\phi_{\rm k}$ & $[0, 2\pi]$ & Uniform $p(\phi) = 1/2\pi$   \\
{SN} mean anomaly of the orbit & $[0, 2\pi]$ & Uniformly distributed  \\
CCSN remnant mass prescription  & delayed &  From \citet{Fryer2012}, which  has no lower {BH} mass gap  \\
USSN remnant mass prescription & delayed &  From \cite{Fryer2012} \\
ECSN  remnant mass presciption & $m_{\rm{f}} = 1.26\,M_\odot$ & Based on Equation (8) in \citet{Timmes1996}          \\
CCSN velocity dispersion $\sigma_{\rm{rms}}$ & $100,\,265\,{\rm km}\,{\rm s}^{-1}$ & 1D rms value based on \cite{Hobbs2005} and \cite{Atri2019} \\
USSN and ECSN velocity dispersion $\sigma_{\rm{rms}}$ & $30\,{\rm km}\,{\rm s}^{-1}$ & 1D rms value based on e.g., \cite{Pfahl2002,Podsiadlowski2004}    \\
PISN/PPISN remnant mass prescription & \cite{Marchant2019} & As implemented in \cite{Marchant2019}      \\
Maximum NS mass & $M_{\rm NS,max}=2.2\,M_\odot$ & Based on \cite{Antoniadis2013,Alsing2018,Romani2022}            \\
Tides and rotation & & We do not include prescriptions for tides and/or rotation\\
\hline
\multicolumn{3}{c}{Simulation settings} \\ 
\hline
Total number of binaries sampled per metallicity  & $10^6$ & A million binaries per $Z_{\rm{i}}$ grid point are simulated\\  
Binary fraction & $f_{\rm{bin}} = 1$ & \\
Solar metallicity $Z_\odot$ & $Z_\odot$ = 0.0142 & Based on {\cite{Asplund2009}} \\
Binary population synthesis code                                      & \texttt{COMPAS} (v02.39.00) &  \cite{Stevenson2017,Vigna2018,Neijssel2019} \\
& & \cite{Broekgaarden2019,TeamCOMPAS2022} \\
\hline \hline
\end{tabular}
}
\end{table*}

\bibliography{ms}{}
\bibliographystyle{aasjournal}
\end{document}